\documentclass[10pt]{article}
\usepackage[OE]{express}

\begin{document}

\title{Corrective Re-gridding Techniques for Non-Uniform Sampling in Time Domain Terahertz Spectroscopy}

\author{A.M. Potts,\authormark{1,2,3} T.T. Mai,\authormark{1} M.T. Warren,\authormark{1} and R. Vald\'es Aguilar\authormark{1,*}}

\address{\authormark{1}Department of Physics, The Ohio State University, 191 W Woodruff Ave, Columbus, OH 43210, USA\\
\authormark{2}Department of Electrical and Computer Engineering, The Ohio State University, 2015 Neil Avenue, Columbus, OH 43210, USA\\
\authormark{3}Lake Shore Cryotronics, 575 McCorkle Blvd, Westerville, OH 43082, USA\\}

\email{\authormark{*} valdesaguilar.1@osu.edu} 



\begin{abstract}
Time domain terahertz spectroscopy typically uses mechanical delay stages that inherently suffer from non-uniform sampling positions. We review, simulate, and experimentally test the ability of corrective cubic spline and Shannon re-gridding algorithms to mitigate the inherent sampling position noise. We present simulations and experimental results that show re-gridding algorithms can increase the signal to noise ratio within the frequency range of 100 GHz to 2 THz. We also predict that re-gridding corrections will become increasingly important to both spectroscopy and imaging as THz technology continues to improve and higher frequencies become experimentally accessible.
\end{abstract}


\bibliographystyle{osajnl}
\bibliography{Regridding}

\begin{thebibliography}{10}
\newcommand{\enquote}[1]{``#1''}

\bibitem{dexheimerthz}
S.~L. Dexheimer, \emph{Terahertz Spectroscopy. Principles and Applications}
  (CRC Press, 2007).

\bibitem{YSLeeTHz}
Y.-S. Lee, \emph{Principles of Terahertz Science and Technology} (Springer,
  2009).

\bibitem{Bilbro2012}
L.~Bilbro, \enquote{{Fluctuations of Superconductivity in La2-xSrxCuO4 measured
  with Terahertz Time-domain Spectroscopy},} Ph.D. thesis, The Johns Hopkins
  University (2012).

\bibitem{Naftaly2007}
M.~Naftaly and R.~E. Miles, \enquote{{Terahertz Time-Domain Spectroscopy for
  Material Characterization},} Proceedings of the IEEE \textbf{95}, 1658--1665
  (2007).

\bibitem{Withayachumnankul2014}
W.~Withayachumnankul and M.~Naftaly, \enquote{{Fundamentals of measurement in
  terahertz time-domain spectroscopy},}  (2014).

\bibitem{Basov2011}
D.~N. Basov, R.~D. Averitt, D.~{Van Der Marel}, M.~Dressel, and K.~Haule,
  \enquote{{Electrodynamics of correlated electron materials},} Reviews of
  Modern Physics \textbf{83}, 471--541 (2011).

\bibitem{DSP-Mulgrew}
B.~Mulgrew, P.~Grant, and J.~Thompson, \emph{{Digital Signal Processing:
  Concepts and Applications}} (2002), 2nd ed.

\bibitem{Vieweg2014}
N.~Vieweg, F.~Rettich, A.~Deninger, H.~Roehle, R.~Dietz, T.~G{\"{o}}bel, and
  M.~Schell, \enquote{{Terahertz-time domain spectrometer with 90 dB peak
  dynamic range},} Journal of Infrared, Millimeter, and Terahertz Waves
  \textbf{35}, 823--832 (2014).

\bibitem{Withayachumnankul2008}
W.~Withayachumnankul, B.~M. Fischer, H.~Lin, and D.~Abbott,
  \enquote{{Uncertainty in terahertz time-domain spectroscopy measurement},}
  Journal of the Optical Society of America B \textbf{25}, 1059 (2008).

\bibitem{Mickan2000}
S.~Mickan, D.~Abbott, J.~Munch, X.-C. Zhang, and T.~van Doorn,
  \enquote{{Analysis of system trade-offs for terahertz imaging},}
  Microelectronics Journal \textbf{31}, 503--514 (2000).

\bibitem{Jahn2016}
D.~Jahn, S.~Lippert, M.~Bisi, L.~Oberto, J.~C. Balzer, and M.~Koch,
  \enquote{{On the Influence of Delay Line Uncertainty in THz Time-Domain
  Spectroscopy},} Journal of Infrared, Millimeter, and Terahertz Waves
  \textbf{37}, 605--613 (2016).

\bibitem{Soltani2014}
A.~Soltani, T.~Probst, S.~F. Busch, M.~Schwerdtfeger, E.~Castro-Camus, and
  M.~Koch, \enquote{{Error from delay drift in terahertz attenuated total
  reflection spectroscopy},} Journal of Infrared, Millimeter, and Terahertz
  Waves \textbf{35}, 468--477 (2014).

\bibitem{Letosa1996}
J.~Letosa, M.~Garcia-Gracia, J.~Fornies-Marquina, J.~M. Artacho, J.~Letosa,
  M.~Garcia-Gracia, J.~Fornies-Marquina, and J.~M. Artacho,
  \enquote{{Performance limits in TDR technique by Monte Carlo simulation},}
  IEEE Transactions on Magnetics \textbf{32}, 4--7 (1996).

\bibitem{Proppert2014}
S.~Proppert, S.~Wolter, T.~Holm, T.~Klein, S.~van~de Linde, and M.~Sauer,
  \enquote{{Cubic B-spline calibration for 3D super-resolution measurements
  using astigmatic imaging.}} Optics express \textbf{22}, 10304--16 (2014).

\bibitem{R.1980}
J.~R. and C.~de~Boor, \enquote{{A Practical Guide to Splines.}} Mathematics of
  Computation \textbf{34}, 325 (1980).

\bibitem{Young2017}
T.~Young and M.~J. Mohlenkamp, \emph{{Introduction to Numerical Methods and
  Matlab Programming for Engineers}} (Athens, 2017), 7th ed.

\bibitem{Shannon1949}
C.~E. Shannon, \enquote{{Communication in the Presence of Noise},} Proceedings
  of the IRE \textbf{37}, 10--21 (1949).

\bibitem{Strauch1999}
D.~Strauch and B.~Dorner, \enquote{{Phonon dispersion in GaAs},} Journal of
  Physics: Condensed Matter \textbf{2}, 1457--1474 (1999).

\bibitem{Brown1992}
R.~G. Brown and P.~Hwang, \emph{{Introduction to Random Signals and Applied
  Kalman Filtering}} (John Wiley {\&} Sons, Hoboken, New Jersey, 1992), 2nd ed.

\bibitem{VanExter1989}
M.~van Exter, C.~Fattinger, and D.~Grischkowsky, \enquote{{Terahertz
  time-domain spectroscopy of water vapor},} Optics Letters \textbf{14},
  1128--1130 (1989).

\bibitem{Slocum2015}
D.~M. Slocum, R.~H. Giles, and T.~M. Goyette, \enquote{{High-resolution water
  vapor spectrum and line shape analysis in the Terahertz region},} Journal of
  Quantitative Spectroscopy and Radiative Transfer \textbf{159}, 69--79 (2015).

\bibitem{Slocum2014}
D.~M. Slocum, T.~M. Goyette, and R.~H. Giles, \enquote{{High-resolution
  terahertz atmospheric water vapor continuum measurements},} Proc SPIE
  \textbf{9102}, 91020E (2014).

\bibitem{Withayachumnankul2008a}
W.~Withayachumnankul, B.~M. Fischer, and D.~Abbott, \enquote{{Numerical removal
  of water vapour effects from terahertz time-domain spectroscopy
  measurements},} Proceedings of the Royal Society A \textbf{464}, 2435--2456
  (2008).

\bibitem{Klatt2009}
G.~Klatt, R.~Gebs, C.~Janke, T.~Dekorsy, and A.~Bartels,
  \enquote{{Rapid-scanning terahertz precision spectrometer with more than 6
  THz spectral coverage},} Optics Express \textbf{17}, 22847 (2009).

\bibitem{Mittleman1996}
D.~M. Mittleman, R.~H. Jacobsen, and M.~C. Nuss, \enquote{{T-ray imaging},}
  IEEE Journal on Selected Topics in Quantum Electronics \textbf{2}, 679--692
  (1996).

\bibitem{Zhang2002}
X.~C. Zhang, \enquote{{Terahertz wave imaging: horizons and hurdles.}} Physics
  in medicine and biology \textbf{47}, 3667--3677 (2002).

\bibitem{Mittleman1999}
D.~M. Mittleman, M.~Gupta, R.~Neelamani, R.~G. Baraniuk, J.~V. Rudd, and
  M.~Koch, \enquote{{Recent advances in terahertz imaging},} Applied Physics B:
  Lasers and Optics \textbf{68}, 1085--1094 (1999).

\bibitem{Huang2009}
S.~Huang, P.~C. Ashworth, K.~W. Kan, Y.~Chen, V.~P. Wallace, Y.-t. Zhang, and
  E.~Pickwell-MacPherson, \enquote{{Improved sample characterization in
  terahertz reflection imaging and spectroscopy},} Optics Express \textbf{17},
  3848 (2009).

\bibitem{Siegel2004}
P.~H. Siegel, \enquote{{Terahertz technology in biology and medicine},} IEEE
  Transactions on Microwave Theory and Techniques \textbf{52}, 2438--2447
  (2004).

\bibitem{Woodward2002}
R.~M. Woodward, B.~E. Cole, V.~P. Wallace, R.~J. Pye, D.~D. Arnone, E.~H.
  Linfield, and M.~Pepper, \enquote{{Terahertz pulse imaging in reflection
  geometry of human skin cancer and skin tissue},} Physics in Medicine and
  Biology \textbf{47}, 3853--3863 (2002).

\bibitem{HernandezCardoso2017}
G.~Hernandez-Cardoso, S.~Rojas-Landeros, M.~Alfaro-Gomez, A.~Hernandez-Serrano,
  I.~Salas-Gutierrez, E.~Lemus-Bedolla, A.~Castillo-Guzman, H.~Lopez-Lemus, and
  E.~Castro-Camus, \enquote{{Terahertz imaging for early screening of diabetic
  foot syndrome: A proof of concept},} Scientific Reports \textbf{7}, 42124
  (2017).

\bibitem{Zimdars2004}
D.~A. Zimdars and J.~S. White, \enquote{{Terahertz reflection imaging for
  package and personnel inspection},} in \enquote{Proc. SPIE,} , vol. 5411
  (2004), vol. 5411, p.~78.

\end{thebibliography}

\section{Overview of Time Domain Terahertz Spectroscopy}

Typical time domain terahertz spectrometers (TDTS) are quasi-optical instruments used to measure the response of a sample to a broadband electromagnetic pulse whose frequency components dominantly lie between 100 GHz and 2.0 THz  \cite{dexheimerthz,YSLeeTHz,Bilbro2012}. Generating a THz pulse starts with an ultrafast laser, that ideally, generates a Gaussian pulse with a temporal pulse width on the order of a few tens of femtoseconds \cite{dexheimerthz,YSLeeTHz}. For TDTS based on photoconductive sources, the femtosecond pulse is focused onto an electrically-biased emitter antenna. Transient carrier dynamics produce a broadband output pulse with most of the energy contained at frequencies less than 2.0 THz \cite{dexheimerthz,YSLeeTHz,Naftaly2007,Withayachumnankul2014}, giving researchers a tool to probe many emergent condensed matter phenomena, such as superconducting energy gaps, Josephson plasmons, and antiferromagnetic magnons at their natural energy scales \cite{Basov2011}.

The width of the main time domain peak of a THz pulse is on the order of one picosecond. Such small times and broadband excitations make conventional electrical sampling techniques, such as high-speed oscilloscopes and bandpass sampling, of little use. An elegant solution is to use another antenna as a detector and a mechanically-controlled optical delay stage to sample the THz electric field as a function of position. The femtosecond laser pulse responsible for generating the THz pulse is split into two beams. One beam bounces off of a retroreflector on an mechanical delay stage and reaches an emitter antenna, which produces THz radiation that propagates through the sample and reaches the detector antenna. This THz electric field creates a potential difference across the detector antenna and induces a current that flows only when the antenna has been turned on (\textit{gated}) by a laser pulse. By varying the optical delay stage position, we can vary the optical path length difference between the THz pulse and pulse responsible for gating the detector. Sweeping the delay stage through positions corresponding to about 70 picoseconds in time allows us to sample the entire THz waveform directly in the time domain. Applying an AC voltage bias on the emitter antenna and using a lock-in amplifier at the detector output greatly reduces broadband systematic noise \cite{Naftaly2007,Withayachumnankul2014}. The current amplifier and lock-in detector operate at a frequency on the order of one kilohertz. Given a laser pulse repetition rate of 80 MHz, each complete period on the lock-in amplifier corresponds to averaging tens of thousands of detected current pulses to produce a DC signal. The resulting current is then converted to a voltage with a low noise transimpedance amplifier that feeds to a lock-in amplifier. This signal is then recorded on a computer. 

In order to determine the optical constants of a sample under consideration, a complete measurement requires two scans. In a transmission TDTS, the first scan is the reference and contains no sample. The other scan is recorded with the sample placed in the THz beam path. The scans are then multiplied by a window function which ensures the first and last point have zero value. This is done for accurate amplitude and spectral resolution \cite{DSP-Mulgrew}. The ratio of the Fourier transform of the sample output to the Fourier transform of the reference scan gives the sample transmission function in the frequency domain  \cite{Naftaly2007,Withayachumnankul2014}. Having a large dynamic range, therefore, is imperative to accurately obtain the sample\textquotesingle s optical constants as a function of frequency \cite{Naftaly2007,Withayachumnankul2014,Vieweg2014}.

\section{Sampling Techniques and Corrective Algorithms}

The precision of a TDTS system\textquotesingle s output waveform is a function of the laser's stability, noise in antennas, and the accuracy and precision of delay stage\textquotesingle s sampling position resolution \cite{Withayachumnankul2014, Withayachumnankul2008, Mickan2000}, amongst other sources. Delay stages cannot instantly move to a sampling position with infinite precision; there is a fundamental trade-off between sampling position precision and measurement time. Some degree of sampling position precision must be sacrificed in order to complete a measurement in a reasonable amount of time. The effects of sampling position error are magnified when the sample transmission function is computed by taking a ratio of the fast Fourier transform (FFT) of the sample scan to that of the reference scan, as both scans have corresponding data points taken at different sampling positions. These deviations from regular sampling positions create noise in the computation of the sample\textquotesingle s transmission function, and thus of its optical constants \cite{Withayachumnankul2008}. 

Smaller sampling position tolerances cause scans to take longer amounts of time, as the closed loop control process in the delay stage takes more time to find a more precise position. One simple solution to irregular sampling positions is to decrease the delay stage tolerance and take a scan over a large amount of time \cite{Jahn2016}. This is inconvenient for the user, but more importantly may cause a fundamental assumption to break down. The preceding discussion assumes that all transmitted pulses are identical, which is expected to be true for TDTS based on photoconductive switches. This is because the repeatability of the pulses only depends on the statistical process of carrier recombination in the semiconducting material. However, small long-term drifts in pump laser power have been observed on a timescale of one hour, even in well-controlled, mode-locked femtosecond lasers. There are also other long term drifts in the measurement environment, such as variations in temperature and humidity, which can affect the quality of long measurements\cite{Withayachumnankul2008, Jahn2016, Soltani2014}.
 
Signal processing techniques can be used to mathematically understand and correct for delay stage imprecision \cite{Letosa1996}. We propose, simulate, measure, and compare the effectiveness of several signal processing techniques in order to compensate for irregular sampling positions without fundamentally changing the measurement setup or increasing the time of the scan. All corrective techniques use different algorithms to compute the best estimate of the waveform at uniformly spaced positions. The Fourier transform of this re-gridded waveform is then computed and analysis proceeds as if there were no sampling position precision error. Note that each algorithm must be applied twice: one to the sample scan and once to the reference scan. We will theoretically and experimentally explore three such algorithms: conventional analysis, cubic spline, and Shannon re-gridding.

\subsection{Conventional Analysis}
Conventional analysis in TDTS assumes that all sampling positions are regularly spaced. The data is converted to the frequency domain via an FFT and conventional analysis proceeds. Conventional analysis works very well if the delay stage's sampling positions deviate very little from their ideally uniform values. This is true in experimental setups with high-quality delay stages. High-end optical delay stages, operating with closed loop control, often have typical sampling position deviations of no more than 10\% with typical deviations of about 1\% or less, generally making conventional analysis feasible for processing data. Good delay stages, however, are expensive and may operate slowly with closed loop control turned on. 

\subsection{Cubic Spline Re-gridding}
\begin{figure}[b]
\centering\includegraphics[width=9cm]{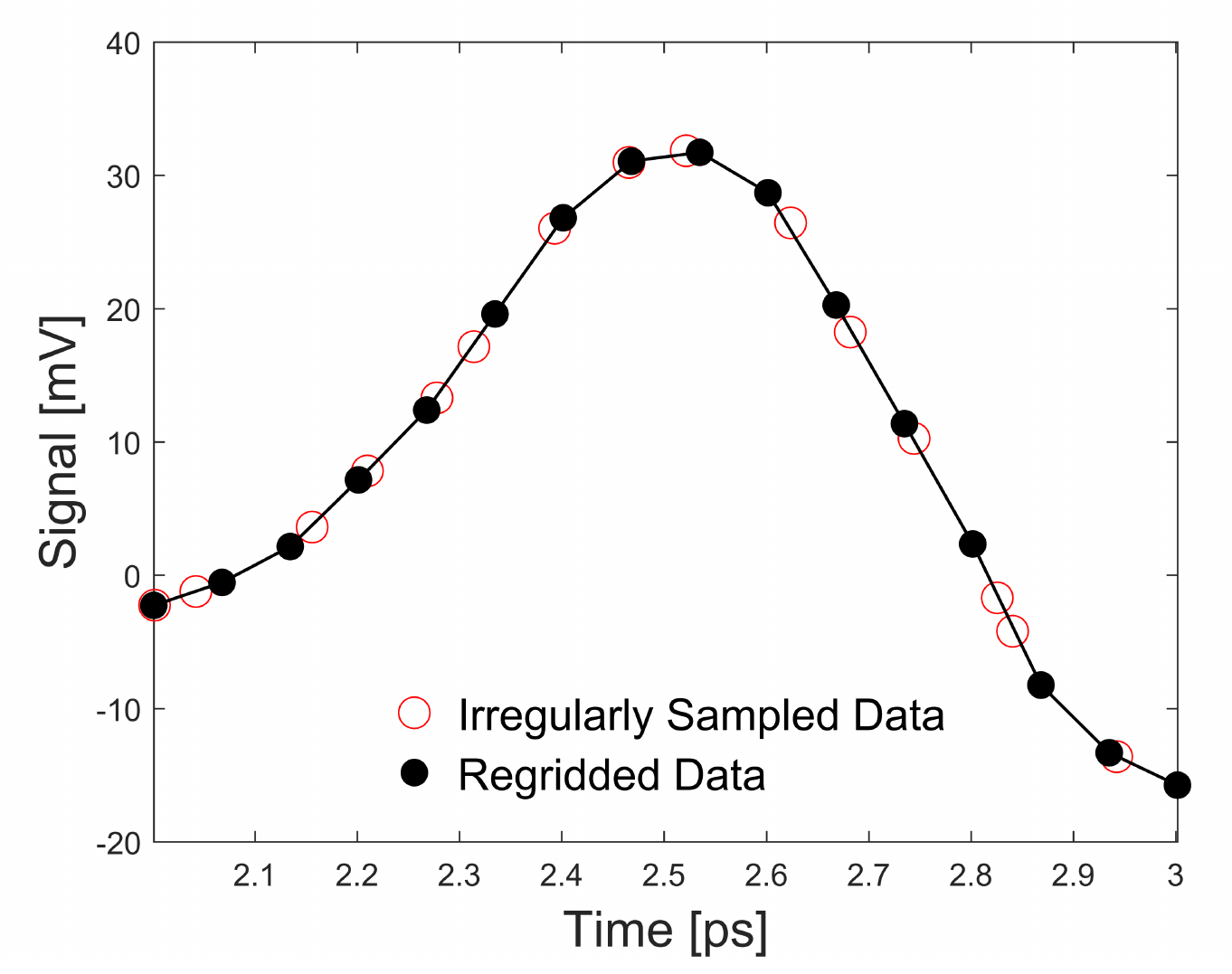}
\caption{Illustration of the re-gridding process of the main peak for a real THz pulse. This process takes irregularly sampled data points (red) and produces a regularly sampled estimate of the pulse (black). The line is a guide to the eye.} 
\label{timeDomain}
\end{figure}

Cubic spline re-gridding, a technique commonly used in ultrafast optics \cite{Proppert2014}, is used to interpolate between irregularly sampled points in time. Each local interval between two points is assigned a local and unique cubic interpolating polynomial, which represents a continuous estimate of the original analog waveform\textquotesingle s value. The piecewise cubic spline is evaluated at desired regular sampling points to generate an estimate of the original waveform\textquotesingle s value. The cubic spline thus generates continuous function that may be evaluated at any time domain point, whether it was physically sampled or not. An illustration of this re-gridding process with actual data may be seen in figure \ref{timeDomain}. 

The proper generation of the cubic polynomial coefficients for each interval is well-studied \cite{R.1980}. Generating these coefficients on the interior of the set of sampled points (where edge effects are irrelevant) is done as follows \cite{Young2017}. Any two neighboring cubic splines $S_k$ and $S_{k+1}$ are defined by:

\begin{equation}
S_k (x)=A_k+B_k x_k+C_k x_k^2+D_k x_k^3
\end{equation}
\begin{equation}
S_{k+1} (x)=A_{k+1}+B_{k+1} x_{k+1}+C_{k+1} x_{k+1}^2+D_{k+1} x_{k+1}^3\,.
\end{equation}

The interpolating procedure requires us to uniquely resolve every cubic spline by solving for the coefficients $A_k, B_k, C_k,$ and $D_k$ for all values of k. The cubic splines defined at the $k^{th}$ and $k+1^{th}$ points ($S_k$ and $S_{k+1}$, respectively)  are required to go through the y-values sampled at $x_k$ and $x_{k+1}$. Namely, we set $S_k(x_k) = y_k$ and $S_{k+1}(x_{k+1}) = y_{k+1}$. The first and second derivatives of the $k^{th}$ spline are then matched to the first and second derivatives of the $k+1^{th}$ spline at the sampled points $x_k$ and $x_{k+1}$, yielding additional independent equations for $A_k, A_{k+1}, B_k, B_{k+1} \hdots$ etc. This procedure produces four linear equations, allowing us to solve for $A_k, B_k, C_k$ and $D_k$ for each local cubic spline. Resolving all coefficients of the cubic splines bordering the zeroth and final elements requires two additional equations. We then set the second derivatives of the cubic splines at the endpoints to zero, so-called natural boundary conditions \cite{R.1980,Young2017}.

\subsection{Modified Shannon Interpolation}
We propose a mathematical mechanism to accurately compensate for non-uniform sampling positions based on Shannon interpolation \cite{Shannon1949}. Typical photoconductive antennas down-convert most of the energy in the pump laser pulse into frequencies between 100 GHz and 2.0 THz. The low optical phonon in the GaAs emitter and detector antennas is known to strongly attenuate frequencies of a few terahertz and above \cite{Strauch1999}. One can therefore assume, with small error, that the time domain terahertz waveform is broadband but band-limited. 

Shannon interpolation, a famous algorithm used to perfectly reconstruct band-limited signals \cite{Shannon1949}, may be adapted for this situation. Suppose a signal is sampled at irregular sampling positions $x_n$ with output values of y($x_n$). This signal is desired to be known at uniform positions nX with output values of Y(nX). Assuming that the desired, regularly sampled signal Y is band-limited allows us to apply Shannon\textquotesingle s Theorem and consider the irregular signal y to be an ideal interpolation of Y. This, mathematically, is written as:

\begin{equation}
\label{sincMatrix}
\begin{pmatrix}
y(x_1) \\
y(x_2) \\
\vdots \\
y(x_n) \\
\end{pmatrix}
= 
\begin{pmatrix}
sinc\left( \frac{x_1 - X}{X} \right) & sinc\left(\frac{x_1 - 2X}{X}\right) & \hdots & sinc\left( \frac{x_1 - nX}{X} \right) \\
sinc\left( \frac{x_2 - X}{X} \right) & sinc\left( \frac{x_2 - 2X}{X} \right) & \hdots & sinc\left( \frac{x_2 - nX}{X} \right) \\
\hdots & \hdots & \ddots & \vdots \\
sinc\left(\frac{x_n - X}{X} \right) & sinc\left( \frac{x_n - 2X}{X}\right) & \hdots & sinc\left( \frac{x_n - nX}{X} \right) \\
\end{pmatrix}
\begin{pmatrix}
Y(X) \\
Y(2X) \\
\vdots \\
Y(nX) \\
\end{pmatrix}
\end{equation}

The $sinc$ function is defined by $sinc(x)=\frac{\sin(\pi x)}{\pi x}$. The matrix equation may be written more concisely as:
\begin{equation}
\boldsymbol{y}=\overline{R}\boldsymbol{Y}
\end{equation}

We note that the matrix reduces to the identity matrix in the case of uniform sampling. Given the regular sampling locations contained in the data set \textbf{y}, one can solve for \textbf{Y} by evaluating:

\begin{equation}
\boldsymbol{Y} = \overline{R}^{\:-1} \boldsymbol{y}
\end{equation}
$\overline{R}^{\:-1}$ matrices of rank N are difficult to be manipulated analytically, and computing this inverse numerically can be computationally intense for very large data sets, but it is somewhat trivial for a few thousand points or less, which is the typical size of TDTS scanned data. The resulting interpolated waveform \textbf{Y} is then used to determine the transmission function as described above. The resulting reconstructed electric field pulses are only sampled at the ideal sampling positions; this algorithm does not generate continuous interpolating function.

\subsection{Non-Causal Re-gridding}

Examination of the matrix in equation \ref{sincMatrix} reveals that, in some sense, the Shannon algorithm appears to be non-causal. Consider some re-gridded output value $Y_m$ evaluated after data taken at irregular sampling position. It is clear that $Y_m$ is a linear combination of {$\hdots y_{m-1},y_m,y_{m+1}\hdots$}, whose coefficients come from $\overline{R}^{\:-1}y_m$. Thus, $Y_m$ depends on both past (such as $y_{m-1}$), the present ($y_m$), and future ($y_{m+1}$) input values. The physical meaning of using non-causal re-gridding techniques is perplexing at first inspection. Physically, the present value of the electric field cannot depend on the value of any input parameters at any time in the future. Equation \ref{sincMatrix}, however, seems to suggest that the electric field at present time depends on future values of input. 

The resolution to such a concern requires careful examination of these equations and of the physical meaning of causality. During measurement, the present value of an electric field does not depend on future values of electric field, i.e. causality applies. After data collection is complete, re-gridding techniques are used to find a best \textit{estimate} of the waveform at a new point between two known points. The $n^{th}$ sampled output value of the estimate does depend on the $n+1^{th}$ point, but the $n^{th}$ point of electric field does not depend on $n+1^{th}$ point of the electric field. Re-gridding techniques do not demand that present physical quantities depend on future values of those physical quantities: re-gridding techniques only assume that the relevant physical quantity has mathematical support between the known points. 

Most analysis techniques appear, by their very nature, to be non-causal. Computing a numerical derivative at a point $x = n$ using central differences requires knowing the value of the future point $y_{n+1}$. Least-squares fitting algorithms use all values of a waveform to find a best estimate of desired parameters. Non-causal Weiner filters are commonly used in analysis of avionic system data  \cite{Brown1992}. It can be shown that non-causal Weiner filters, namely Weiner filters that use the data set as a whole during post-processing, produce significantly lower error than their causal cousins \cite{Brown1992}. Ignoring non-causal re-gridding techniques may be equivalent to ignoring low-error estimates.  Non-causality that requires post-processing poses no threat to altering the causal physics captured by the data, Shannon and spline re-gridding techniques although non-causal-looking, seem to not alter the physical relation between \textit{real} present electric field values and future ones.

\section{Simulation}

\begin{figure}[t]
\centering\includegraphics[width=13.5cm]{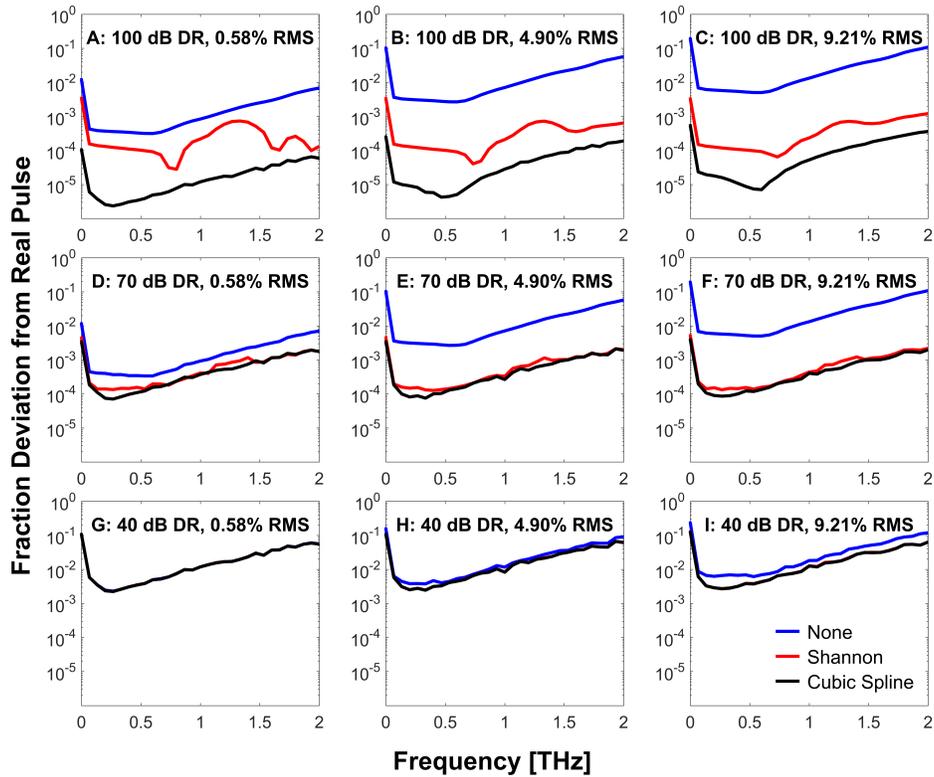}
\caption{Simulated re-gridding error as a function of frequency for varying dynamic range (DR) and different percent RMS sampling errors (RMS) averaged over 100 simulated scans. Re-gridding becomes increasingly effective with increasingly high dynamic range and increasingly large sampling position deviations. Note that the Shannon and cubic spline terms produce errors indistinguishable from each other on this scale when the dynamic range is 40.0, which is somewhat typical of a TDTS system.}
\label{simData}
\end{figure}

Corrective re-gridding algorithms were numerically simulated with a pulse generated from the model described in \cite{Bilbro2012} with a laser pulse width of 40 fs, an electron-hole recombination time of 500 fs, and a transport scattering time of 400 fs. The pulse was sampled N times at non-uniform sampling positions; non-uniform positions were simulated by sampling the pulse at random points within plus or minus M percent of the regular sampling points, where M ranges from 2\% to 33\%. As in the experiments, such errors do not compound throughout the course of the scan: each sampling point is subject to the same tolerance as the previous and subsequent points. Corrupting white noise was added to the signal y-values to simulate the effects of a finite experimental dynamic range. The performance of re-gridding algorithms is tested in response to varying RMS x-axis (sampling position) errors and y-axis (output waveform) noise. The results of such simulation is shown in figure \ref{simData}.

Figure ~\ref{simData} shows the capability of re-gridding techniques to reduce errors due to irregular sampling positions. Shown is the relative difference between the FFT of a regularly-sampled waveform and the FFT of an irregularly-sampled waveform that has been re-gridded, normalized to the regularly-sampled waveform, giving a normalized measurement of the error associated with the re-gridding process. The ability of these algorithms to reduce sampling position noise is strongly dependent on the dynamic range, i.e. the accuracy of the measurement of "y" values. If the dynamic range were infinite, the ideal (Shannon) re-gridding algorithm would produce an estimate whose accuracy is limited only by the assumption of a purely bandlimited signal \cite{Shannon1949}. Re-gridding offers minor improvement to correct for sampling position imprecision when the signal dynamic range is only two orders of magnitude (40 dB). When the dynamic range is increased to 3.5 orders of magnitude (70 dB), both the cubic spline and Shannon  techniques become noticeably more successful in mitigating noise. The associated error substantially decreases when the dynamic range is increased to 5 orders of magnitude (100 dB). As the errors in the sampling position increase (measured here as an RMS percent of the total step size), the re-gridding errors increase, especially at high dynamic ranges (panels A - C). We note that the simulation does indicate that the cubic spline algorithm outperforms the Shannon algorithm at extremely high dynamic ranges. This could be a result of the Shannon algorithm\textquotesingle s assumption of a bandlimited signal.

\section{Experimental Testing}
The home-built TDTS system at OSU, which uses a lock-in amplifier with a two pole filter and 100 ms time constant, can achieve a dynamic range of about 40 - 45 dB when collecting one single data point per sampling position. Increasing the system\textquotesingle s dynamic range by a few dB may be done by averaging ten to twenty data points per sampling position. Further averaging, however, does little to mitigate y-axis noise. Assuming that the measured y-noise is normally distributed, the system dynamic range (in linear units) increases with the square root of the number of data points. Several hundred data points per sampling position are required to significantly elevate the system\textquotesingle s dynamic range. Given the aforementioned objective of reducing sampling time to avoid systematic, long-term drifts by mitigating sampling position errors, excessive averaging is not desirable.

The well-documented absorption of THz radiation by water vapor offers a convenient and universal spectrum for calibrating TDTS measurements \cite{VanExter1989, Slocum2015, Slocum2014, Withayachumnankul2008a, Klatt2009}. The spectrometer\textquotesingle s measurement chamber is first purged with Nitrogen gas to remove water vapor. Several scans were taken and averaged to generate an average reference spectrum. The spectrometer\textquotesingle s measurement chamber was then allowed to fill with air with high water vapor content. The exact absolute humidity was not known, but the relative humidity was $\approx$45\% at a temperature of 72$^\circ$ Fahrenheit. Re-gridding algorithms are applied to each trace in the time domain. We recover the transmission function of the water vapor alone by referencing the humid spectrum to the purged spectrum \cite{Withayachumnankul2014}. 

\begin{figure}[t]
\centering\includegraphics[width=12cm]{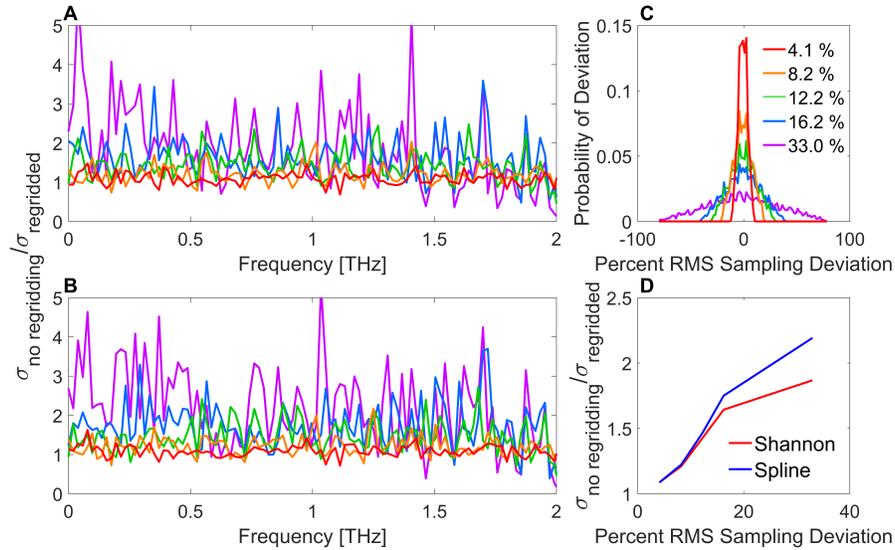}
\caption{Improvement versus frequency (for experimental data) by re-gridding process for Shannon (panel A), and cubic spline re-gridding (panel B) with various artificially-generated sampling position error distributions with the RMS error quoted in panel C. Measurement steps in the delay stage of 10 microns at a dynamic range of approximately 40 dB were used. The improvement in the standard deviation by re-gridding averaged from DC to 2.0 THz versus RMS sampling deviations (panel D) shows ratios greater than unity. Cubic spline is consistently more effective than Shannon re-gridding at high sampling position deviations.}
\label{regriddingComparison}
\end{figure}

Closed loop control is largely responsible for moving the delay stage as close to the desired sampling position as possible, but it does take a long time to operate (compared to the rest of the sampling process). By artificially sending the delay stage to a random position with a deviation between two fixed percentages of the average change in sampling position per point, we imitate a lower-end commercial delay stage capable of sampling with very limited position control. Effectively, we imitate a delay stage operating with nonfunctional closed loop control. We take fifteen full scans per allowed RMS tolerance. The uncertainty of a particular measurement is evaluated by computing the fast Fourier transform of each scan, referencing to the average re-gridded (purged) scan, and measuring the standard deviation of each point as a function of frequency in the spectral range where the system has an appreciable signal to noise ratio (100 GHz to 2 THz, as seen in figure \ref{regriddingComparison}). We compare the ratio of the standard deviation without any re-gridding to the standard deviation of the traces following the application of the re-gridding algorithm (both the reference and the actual scans). A ratio less than unity means that the re-gridding process increased the experimental uncertainty in each point and is thus detrimental. A ratio of unity means that re-gridding offers no improvement. A ratio greater than unity means that re-gridding has decreased the experimental uncertainty, yielding a more precise estimate of water vapor\textquotesingle s THz transmission function. Results of data collection at a dynamic range of approximately 40 dB may be seen in figure \ref{regriddingComparison}.

Figure \ref{regriddingComparison} shows that re-gridding offers some improvement from 100 GHz to 2 THz when the dynamic range is 40 dB no matter the tolerance, as the average ratio of standard deviation without re-gridding to that with re-gridding is always greater than unity (panel D of figure \ref{regriddingComparison}). The improvement monotonically climbs as the RMS sampling position deviation increases, which is consistent with the trend seen with RMS increased from 0.58\% to 9.21\% at 40 dB shown in panel I of figure \ref{simData}. The improvement is small, but consistent with the order of magnitude predicted by simulation, implying that re-gridding will become more significant as the system\textquotesingle s dynamic range improves. Even though the sampling position distributions in figure \ref{regriddingComparison} are broader than those in a typical high-end delay stage (since the delay stage was forced to go to non-uniform sampling positions) they offer substantial evidence that re-gridding can play a significant role in mitigating noise due to delay line irregularity.

\section{Summary}

Three primary conclusions may be drawn from these computational and experimental studies. (1) The noise in a typical TDTS does not depend strongly on deviations from ideal sampling positions when using closed loop control, and the RMS sampling position deviation is less than 1\% than the sampling step size.  (2) Re-gridding will be an important technique to mitigate sampling position noise at high dynamic ranges and at higher frequencies in TDTS. (3) Re-gridding may allow some systems, particularly in the field of imaging, to maintain a reasonable signal to noise ratio while greatly decreasing measurement acquisition time by using a low-end or fast delay stage.

Noise caused by sampling position error in a delay stage is small in comparison to other sources of y-axis noise when the delay stage operates with closed loop control. Depending on the particular system, these could include noise from photoconductive antennas, thermal drifts and fluctuations in the laser, parasitic reflections, among others \cite{Withayachumnankul2008}. Research should concentrate on lowering noise from such sources before further examining noise from the non-ideal sampling positions.

Once other sources of noise have been minimized, theory and simulations suggest that sampling position imprecision can be mitigated through the signal processing algorithms discussed here. Re-gridding processes show small but nonzero improvement in a modern research-grade TDTS with typical sampling position deviations, but show more significant improvement when the sampling position deviations are exaggerated. Both Shannon and cubic spline algorithms experimentally produce comparable improvements over not re-gridding at all, and this is consistent with simulations. Though presently immeasurable by experiment, our simulations predict that re-gridding will be important to reducing measurement errors for dynamic ranges of at least 70 dB. Simulations predict that the cubic spline will be more effective than Shannon re-gridding at extreme dynamic ranges. Experimental data shows a small improvement with high sampling position errors in the spline compared to the Shannon re-gridding, but further experimental testing at higher dynamic range and higher frequencies is required to more rigorously explore this.

Re-gridding does offer an exciting prospect for quickly taking terahertz scans, particularly in the field of reflection imaging. Though a powerful non-destructive technique for material \cite{Mittleman1996,Zhang2002,Mittleman1999,Huang2009}, biological \cite{Zhang2002,Siegel2004,Woodward2002,HernandezCardoso2017} and security/military \cite{Zimdars2004} applications, THz reflection imaging suffers from the long times required to generate images of relatively small objects, making its use in practical security scanners and imagers difficult to practically implement \cite{Zimdars2004}. The need for precise sampling positions requires closed-loop control, which takes more time than simple open loop control. Even well-designed systems spend a comparatively long time properly moving the delay stage to image each pixel. If the system has a sufficiently high dynamic range, one could take a fast scan without closed loop control and use corrective re-gridding algorithms to quickly generate a THz image. This would remove a substantial amount of time in the sampling process and make rapid THz imaging more feasible to a wider community of scientists and engineers. 

\section*{Acknowledgments}
We thank Peter Armitage, Dipanjan Chaudhuri, Nicholas Crescimanno, David Daughton, Daniel Heligman, Evan Jasper, Ciriyam Jayaprakash, and Jeffrey Lindemuth for useful discussions and suggestions.

\section*{Funding}
Partial support has been provided by the Institute for Materials Research at OSU under grant EMRG-G00030. AMP acknowledges partial support by Lake Shore Cryotronics.

\section*{Disclosures}
The authors declare that there are no conflicts of interest related to this article.

\end{document}